\documentclass[12pt]{article}
\newcommand{\doublespacing}{\baselineskip=26pt plus 0.5pt minus 1pt}
\usepackage[dvips]{graphicx}
\font\scaps=cmcsc10    % small capitals
\newcommand \be{\begin{equation}}
\newcommand \ee{\end{equation}}
\newcommand \ba{\begin{eqnarray}}
\newcommand \ea{\end{eqnarray}}

\begin{document}

 \setkeys{Gin}{draft=false}    % to produce draft

\def\today{\ifcase\month\or
 January\or February\or March\or April\or May\or June\or
 July\or August\or September\or October\or November\or December\fi
 \space\number\day, \number\year}
%
%\hfil PostScript file created: \today{}; \ time \the\time \ minutes
\vskip .05in

% \title{Testing spatial distribution forecast of point
% processes with application to seismology }

\title{Testing long-term earthquake forecasts:
likelihood methods and error diagrams }

\author{Yan Y. Kagan$^1$\footnote{Yan Y. Kagan, Department of
Earth and Space Sciences, University of California, Los
Angeles, California, 90095-1567, USA; (e-mail:
ykagan@ucla.edu)}
\\ $^1$ Department of Earth and Space Sciences\\ University of
California, Los Angeles, California, USA} \maketitle

\begin{abstract}
\doublespacing

We propose a new method to test the effectiveness of a spatial
point process forecast based on a log-likelihood score for
predicted point density and the information gain for events
that actually occurred in the test period.
The method largely avoids simulation use and allows us to
calculate the information score for each event or set of
events as well as the standard error of each forecast.
As the number of predicted events increases, the score
distribution approaches the Gaussian law.
The degree of its similarity to the Gaussian distribution can
be measured by the computed coefficients of skewness and
kurtosis.
To display the forecasted point density and the point events,
we use an event concentration diagram or a variant of the
{\sl Error Diagram} (ED).

We demonstrate the application of the method by using our
long-term forecast of seismicity in two western Pacific
regions.
We compare the ED for these regions with simplified diagrams
based on two-segment approximations.
Since the earthquakes in these regions are concentrated in
narrow subduction belts, using the forecast density as a
template or baseline for the ED is a more convenient display
technique.
We also show, using simulated event occurrence, that some
proposed criteria for measuring forecast effectiveness at EDs
would be strongly biased for a small event number.

{\scaps {Key words}: }
Probabilistic forecasting;
Spatial analysis;
Fractals and multifractals;
Probability distributions;
Earthquake interaction, forecasting, and prediction;
Statistical seismology.
% Seismicity and tectonics;
% Theoretical seismology;
% Likelihood methods, Information score, Error diagram, forecast
% effectiveness, long-term earthquake forecast

\end{abstract}

%\begin{article}
\doublespacing
% \clearpage
\newpage

\section{Introduction}
\label{YYK_intro}

This paper continues our analysis of stochastic point
process forecast verification (Kagan 2007b).
There ({\it ibid.}) we had discussed two interrelated
methods for measuring the effectiveness of earthquake
prediction algorithms: the information score based on the
likelihood ratio (Kagan 1991) and the ``Error Diagram" (ED).
These methods have been applied (Kagan 2007b) to temporal
renewal stochastic processes, but only for very long processes
with the number of events approaching infinity.

In this work we extend our analysis by
\hfil\break
1) discussing spatial (not temporal) random processes
(fields);
\hfil\break
2) considering forecast testing if the number of events is
relatively small;
\hfil\break
3) applying newly developed techniques to long-term earthquake
forecasts.

Two issues are related to the problem of testing point
process forecasts:
\hfil\break
$\bullet$
1) Spatial random point fields density evaluation
and its prediction is a mature discipline with many
publications.
Baddeley {\it et al.}\ (2005), Baddeley (2007), Daley and
Vere-Jones (2003, 2008) provide reviews.
As we explain below, the earthquake forecasting problem is
different in many respects from regular density evaluation
and requires special treatment.
However, some results of this paper can be applied to test
the forecast of a random spatial pattern.
\hfil\break
$\bullet$
2) Well-developed application methods exist in weather and
climate prediction; their reviews have been recently published
by Jolliffe \& Stephenson (2003), Palmer \& Hagedorn (2006),
DelSole \& Tippett (2007).
These prediction methods consider continuous processes and
fields; however, with necessary modifications some of these
methods can be used for stochastic point processes.

Our main focus here is on the two most widely used
approaches to assessing earthquake prediction methods
(Zaliapin \& Molchan 2004).
Both approaches evaluate how a prediction method reveals new
information about impending earthquake activity.
The first approach starts by estimating the expected
spatio-temporal distribution of seismicity and uses the
classical likelihood paradigm to evaluate predictive power.
Accordingly, it uses the nomenclature of statistical
estimation.
The second one applies the results by Molchan (1990, 1997;
see also Molchan \& Keilis-Borok 2008) who proposed error
diagrams for measuring prediction efficiency.
The EDs plot the normalized rate of failures-to-predict
($\nu$) versus the normalized time of alarms ($\tau$).
The ED can be considered as a time-dependent analog of the
Neyman-Pearson lemma on making a decision: should we
expect an earthquake within a given spatio-temporal region?
Consequently, it uses the language of hypothesis testing.

The error diagram is related to the {\sl Relative Operating
Characteristic} (ROC) (Swets 1973; Mason 2003, pp.~66-76),
used in signal detection and weather prediction efforts.
In the ROC diagrams the success rate of an event
prediction is compared against the false alarm rate.

Starting with Molchan's (1990) paper, previous EDs were almost
exclusively time-dependent.
We apply the ED to time-independent spatial earthquake
distributions.
In some respects, the earthquake spatial pattern is more
difficult to analyze than the temporal distribution.
In the latter case, we have a reasonable null model (the
uniform in time Poisson process) which can be compared to any
test model.
In the spatial case, the simple model of uniformly distributed
seismicity can hardly serve as an initial approximation; even
large earthquakes (which often can be well-approximated by a
Poisson temporal process) are strongly clustered in space.
This property of seismicity is caused by the fractal nature of
earthquake spatial distribution (Kagan 2007a).
Although in our forecast (Kagan \& Jackson 2000) we use a
projection of earthquake centroids on the Earth surface which
smoothes their spatial distribution (Kagan 2007a), the
spatial distribution still preserves a self-similar fractal
pattern with large parts of the Earth practically aseismic.

Diagrams similar to EDs have been used previously to describe
spatial distribution of seismicity: Rong \& Jackson (2002,
their Fig. 3); Kagan {\it et al.}, (2003, their Fig.~5.3);
Helmstetter {\it et al.}, (2007, their Fig.~4); and Shen {\it
et al.}, (2007, their Figs.~1B, 2B) created spatial
``Concentration Diagrams" to characterize the agreement (or
lack thereof) between the predicted seismicity distribution
and future earthquakes.
These diagrams plot the fraction of the event success rate
(equivalent to $1-\nu$) versus the normalized area ($\tau$),
sorted by probability density.
The sorting is largely analogous to water-level threshold
analysis (Zechar \& Jordan 2008).
These concentration diagrams can easily be converted to EDs by
adding an ascending diagonal and then reflecting the plot in
the line ordinate ($\nu=1/2$).

In principle, such diagrams can be treated like ROC plots
where cells with events are considered as success and empty
cells as false alarms.
However, this interpretation encounters difficulties when the
cells are not infinitesimally small, so some may contain more
than one event.
Moreover, as we show below, for a point process on a sphere it
is difficult to define cells of equal size.
Usual sphere subdivision yields unequal cells larger at the
equator and smaller towards the poles.

Characterizing prediction performance is a major challenge for
ED analysis.
Since prediction results are represented by a function
(curve), it is important to find a simple one-parameter
criterion (a functional) that briefly expresses the efficiency
value.
Several functionals have been proposed, each with some
advantages or disadvantages.
For example, Kagan \& Jackson (2006, their Section~5) show
that two ED trajectories with a very different behavior have
the same ``Sum of Errors" ($\nu+\tau$) value, which often is
proposed as a measure of ED forecast efficiency (Kossobokov
2006; Molchan \& Keilis-Borok 2008).

\section{Long-term earthquake forecasts }
\label{YYK_forecast}

Kagan \& Jackson (1994, 2000) present long-term and short-term
earthquake forecasts in several regions using the CMT catalog
(Ekstr\"om {\it et al.}\ 2005; http://www.globalcmt.org/).
The forecasted earthquake rate is calculated as a spatially
smoothed earthquake distribution.
The spatial kernel used in the smoothing has units of
earthquakes per unit area and time.
In our studies it applies to all shallow (depth less or equal
70~km) earthquakes with moment $M = 10^{17.7}$~Nm (magnitude
5.8) and greater.
The kernel is elongated along the fault-plane, which is
estimated from available focal mechanism solutions.

To take into account the impact of earthquakes outside the
forecasted regions boundaries, we consider events up to
1000~km outside the region window.
The rate density from those earthquakes is added to the
forecast density of `inside' events applying the kernel
estimates.
This additional probability density from outside
events is on average balanced by a contribution `leakage'
from many `insider' earthquakes close to the boundaries.

An important feature of Kagan \& Jackson's (1994) method is a
jack-knife like procedure (Silverman 1986) for testing the
predictive power of the smoothing.
It optimizes the kernel parameters choosing those values which
best predict the second half of a catalogue, using a maximum
likelihood criterion, from the first half.
We argue that because the seismicity pattern exhibits a
long-term clustering (Kagan \& Jackson 1991), such a procedure
is better suited to predict the future earthquake rate.
We also assume on an {\it ad hoc} basis (Kagan \& Jackson 1994,
2000) that the background probability density is uniform over
the whole region and integrates to 1\% of the total earthquake
rate
\be
{\rm (Background \ rate)} = \epsilon \, \times \, {\rm (Total
\ rate)} \, ,
\label{Eq_inf-1}
\ee
with $\epsilon \, = \, 0.01$.

Kagan (2007a) shows that the fractal dimension of earthquake
hypocenters, $\delta$, strongly depends on the earthquake
catalog time interval.
For temporally short catalogs $\delta$ is close to zero and
approaches the asymptotic value $\delta \approx 2.3$ for
catalogs of decades length.
In a more intuitive setting, this result signifies that in
short time intervals, hypocenters are concentrated in a few
point clouds. With increased time, seismicity spreads over a
fault system or seismic belt length, eventually occupying a
set with the dimension in excess of the 2-D plane.
Therefore, if one uses a set of earthquake epicenters in a
relatively short catalog to predict the future seismicity rate,
the optimal forecast kernel should spread beyond the presently
available event points, i.e., to be smoother than the standard
density estimators (Silverman 1986) would suggest.

The forecasts are expressed as the rate density (that is,
the probability per unit area and time).
They are updated every day and posted for two
western Pacific regions at
\hfil\break
http://scec.ess.ucla.edu/$\sim$ykagan/predictions\_index.html
(see FORECAST TEST FOR 2004-2006:).
Table~\ref{tab:Tab1} displays a small extract of the forecast
tables available at the Web site.
In this Table we sorted (ordered) entries by the values of
earthquake forecast densities.

\begin{figure}
\begin{center}
\includegraphics[width=0.575\textwidth,angle=0]{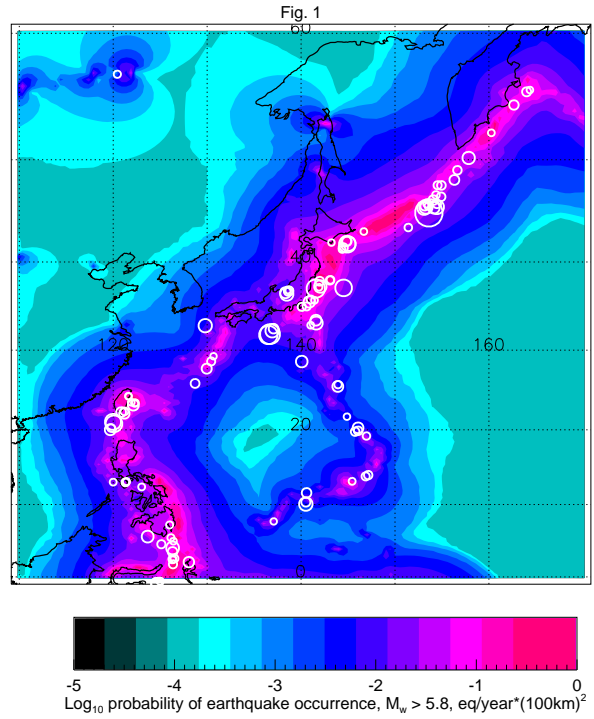}
\caption{\label{yyk_fig0}
NW-Pacific long-term seismicity forecast: latitude
limits from $0.25^{\circ}$S to $60.25^{\circ}$N, longitude
limits from $109.75^{\circ}$E to $170.25^{\circ}$E.
The forecast is calculated at $121 \times 121$ grid.
Colour tones show the probability of shallow
(depth less or equal to 70~km) earthquake
occurrence calculated using the CMT 1977-2003 catalog;
108 earthquakes for 2004-2006 are shown in white.
The uniform background probability density ($\epsilon=0.01$,
see Eq.~\ref{Eq_inf-1}) can be observed at northwest and
southeast corners of the map as greyish-green areas.
}
\end{center}
\end{figure}

In Fig.~\ref{yyk_fig0} we display the long-term forecast map
computed for the northwest (NW) Pacific region using the CMT
catalog for 1977-2003.
Shallow earthquakes in 2004-2006 are shown in white.
Similar maps for NW and southwest (SW) Pacific are shown, for
instance, in Figs.~8a,b by Kagan \& Jackson (2000).
On visual inspection, the model predicts the spatial
distribution of seismic activity reasonably well.
We tested this forecast by a Monte-Carlo simulation (Kagan \&
Jackson 1994).

\begin{figure}
\begin{center}
\includegraphics[width=0.575\textwidth,angle=0]{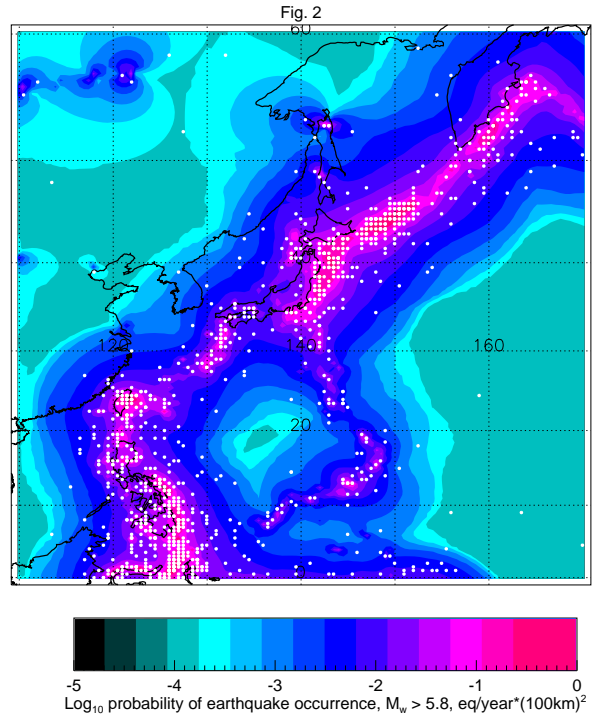}
\caption{\label{yyk_fig1}
NW-Pacific long-term seismicity forecast.
Colour tones show the probability of earthquake occurrence
calculated using the CMT 1977-2003 catalog; 1080 simulated
earthquakes for 2004-2006 are shown in white.
}
\end{center}
\end{figure}

In Fig.~\ref{yyk_fig1} we show the forecasted earthquake
density with 10 sets of synthetic catalogs, each having 108
events.
Earthquakes are assumed to occur at the centers of grid cells;
therefore some of the grid points are occupied by more than
one event.
Some of the simulated points occur in areas of low seismicity
(compare Fig.~\ref{yyk_fig1} with Fig.~\ref{yyk_fig0}).
As mentioned above, this feature of the forecast is used to
prevent surprises, i.e., an occurrence of earthquakes in zones
where no nearby events happened in 1977-2003.

Table~\ref{tab:Tab0} summarizes annual earthquake rates for
both western Pacific regions.
Because events outside the region's boundaries have influence,
the rates calculated through the smoothing procedure and
evaluated by a direct method (dividing the earthquake numbers
by time interval) are close but do not coincide.

\section{Log-likelihood }
\label{YYK_point1}

Kagan \& Knopoff (1977, see also Vere-Jones 1998)
suggested measuring the effectiveness of the earthquake
prediction algorithm by first evaluating the likelihood ratio
to test how well a model approximates an earthquake
occurrence.
In particular, they estimated the information score,
${\hat I}$, per one event by
\be
{\hat I} \, = \, { { \ell -
\ell_0 } \over n } \, = \, { 1 \over n } \sum_{i=1}^n \log_2 {
p_i \over \xi_i } \, ,
\label{Eq_inf0}
\ee
where $\ell - \ell_0$ is the log-likelihood ratio, $n$ is the
number of earthquakes in a catalog, $\log_2$ is used to obtain
the score measured in the Shannon bits of information, $p_i$
is the probability of earthquake occurrence according to a
stochastic model, conditioned by the past:
\be
p_i \, = \, {\rm Prob} \, \{ \, {\rm an\ event\ in\ } \, (t,
\, t+\Delta) \, | \, I(t) \}
\, ,
\label{Eq_inf0b}
\ee
where $I(t)$ is the past history of the process up to the
moment $t$, and $\xi_i$ is a similar probability for the event
occurrence according to the Poisson process.

The Poisson process rate can be calculated by normalizing
the seismicity level in the forecast regions.
Several rates, such as shown in Table~\ref{tab:Tab0}, can be
used in the normalization.
To make our results comparable to the forecast rate density,
we use $\upsilon_1$ values
\be
\xi_i \, = \, { {\pi \, \upsilon_1} \over { 180.0
\times \left [ \, \sin(\theta_u) - \sin(\theta_l) \, \right ]
\, (\phi_u - \phi_l) \times 111.111^2 \times \, 365.25 } }
\, ,
\label{Eq_inf0c}
\ee
where $\upsilon_1$ is the annual rate of earthquakes in each
region in 1977-2003 (Table~\ref{tab:Tab0}), $\theta_u$ and
$\theta_l$ are the upper and lower latitudes, respectively,
$\phi_u$ and $\phi_l$ ditto for longitudes.
% $\Delta T$ is the catalog time interval ($\Delta T =
% 9,861$~days for 1977-2003; 26.9979 = 9,861/365.25).
For the NW-Pacific region $\xi_i \, = \, 2.6289 \times
10^{-9}$ eq/(day $\times$ km$^2$); for the SW-Pacific $\xi_i
\, = \, 3.3479 \times 10^{-9}$ eq/(day $\times$ km$^2$).

Several methods can be used in calculating the information
score for a set of forecasted events.
Using the forecasted rate values ($\lambda_i$ for cell centers
in which earthquakes occurred) we compute
\be
{I_1} \, = \, { 1 \over n } \sum_{j=1}^n \log_2 {
\lambda_i \over \xi_i }
\, ,
\label{Eq_inf1}
\ee
where $n$ is the number of events.

In Eq.~\ref{Eq_inf1} and in derivations below, we assume that
earthquakes in the cells are identically distributed
independent (i.i.d.) events.
The assumed independence may be challenged by the
clustered nature of earthquake occurrence of which
foreshock-mainshock-aftershock sequences are the most clear
example (Kagan \& Knopoff 1977; Kagan 1991).
However,  given the high magnitude (5.8) threshold for the CMT
catalog, the clustering is less pronounced.
The dependent events on average constitute only about 0.2 of
the total seismic activity (Kagan \& Jackson 2000, Eq.~23).
Thus, we expect that earthquake statistical inter-dependence
would have relatively small influence.

As another option, instead of (\ref{Eq_inf1}) we compute the
information score for the actual epicenter (centroid)
locations ($\lambda_k$)
\be
{I_2} \, = \, { 1 \over n } \sum_{k=1}^n \log_2 {
\lambda_k \over \xi_k }
\, .
\label{Eq_inf2}
\ee

In simulated catalogs we generate multiple (${\cal N} =
10,000$) sets of $n_2$ events (Table~\ref{tab:Tab0}) and
calculate the rate for cell centers as the earthquake location
(see Fig.~\ref{yyk_fig1})
\be
{I_3} \, = \, { 1 \over {n_2 } } \sum_{l=1}^{n_2
} \log_2 { \lambda_l \over \xi_l }
\, .
\label{Eq_inf3}
\ee
and
\be
<{I_3}> \, = \, { 1 \over {{\cal N} } } \sum_{\ell=1}^{
{\cal N} } (I_3)_\ell
\, .
\label{Eq_inf3a}
\ee
This method has an advantage in that we do not need to
calculate the rate densities again, as for $I_2$, but instead
use the previously computed forecast tables (as shown in
Table~\ref{tab:Tab1}) to evaluate the scores.

In Fig.~\ref{yyk_fig2} we display the log-likelihood function
distribution differences for the simulation as shown in
Fig.~\ref{yyk_fig1}.
We simulate $n_2$ earthquake locations according to 1977-2003
forecasts for each region.
Each time we calculate the log-likelihood function and subtract
the log-likelihood function value obtained for the real
catalogue in 2004-2006.
Thus, we display the histogram of $I_3 \, - \, I_2$
(Eqs.~\ref{Eq_inf3} and \ref{Eq_inf2}).

\section{Error diagrams }
\label{YYK_point2}

To test the long-term forecast efficiency numerically, we
calculate the concentration diagram.
To make these diagrams, we divide the region into small cells
(0.5 by 0.5 degrees) and estimate the theoretical forecast
rate of earthquakes above the magnitude threshold for each
cell.
We then count the events that actually occurred in each cell,
sort the cells in the decreasing order of the theoretical
rate, and compute the cumulative values of forecast and the
observed earthquake rates (see Table~\ref{tab:Tab1}).
Similar plots have been used in several of our papers
(Kagan {\it et al.}\ 2003; Helmstetter {\it et al.}\ 2007;
Shen {\it et al.}\ 2007).
In effect, these diagrams are equivalent to the error diagrams
(EDs) proposed by Molchan (1990, 2003) and Molchan \& Kagan
(1992).
But in this case we use the normalized spatial area, not time,
as the horizontal axis.

\subsection{Relation between the error diagram and
information score }
\label{YYK_point3}

We illustrate the ED by a sketch in Fig.~\ref{yyk_ede}.
For the spatial point distribution, this example is easier
to construct and explain than for temporal renewal processes
(Kagan 2007b).
The square's diagonal corresponds to the uniform Poisson
distributions of the points in a region, i.e., a random guess
forecast strategy.
As a test example, we assume that the region consists of
three sub-areas, their surfaces $\tau_i$ is 0.1, 0.5, and 0.4
of the total, and the number of events $\nu_i$ is 0.4, 0.5,
and 0.1, in each zone respectively.
The points in these zones are distributed according to the
Poisson spatial process with the density $\nu_i/\tau_i$.
Then, the information score for such a point distribution can
be calculated as (see Eq.~\ref{Eq_inf0})
\ba
I \, = \, \sum_{i=1}^3 \, \nu_i \, \log_2 {
\nu_i \over \tau_i }
\, & = & \,
0.4 \, \log_2 4.0 \, + \,
0.5 \, \log_2 1.0 \, + \,
0.1 \, \log_2 0.25
\nonumber\\
\, & = & \,
0.8 - 0.2
\, = \, 0.6
\, .
\label{Eq_inf4a}
\ea
For the normalized point Poisson distribution in the ED, the
point density is unity.
Hence its contribution to the information rate
(\ref{Eq_inf4a}) is zero.

The information score can be calculated for
continuous concave curves in an error diagram (Kagan 2007b)
\be
I \, = \, \int_0^1 \log_2 \left ( - { { \partial \nu } \over
{ \partial \tau } } \right ) d \nu
\, .
\label{Eq_inf4}
\ee
If the ED consists of several linear segments (as in
Fig.~\ref{yyk_ede}), then (\ref{Eq_inf4}) converts to
\be
I_0 \, = \, \sum_{i=1}^{N} \nu_i \,
\log_2 \left ( { { \nu_i }
\over {\tau_i } } \right )
\, ,
\label{Eq_inf5}
\ee
where $i$ are cell numbers, $N$ is the total number of grid
points, and $\nu_i$ and $\tau_i$ are the
normalized rates of occurrence and cell area:
\be
\nu_i \, = \, { { R_i } \over { \sum_{i=1}^{N} R_i } } \quad
{\rm and } \quad
\tau_i \, = \, { { S_i } \over { \sum_{i=1}^{N} S_i } }
\, ,
\label{Eq_inf5a}
\ee
see Table~\ref{tab:Tab1}.
When such calculations are made for a spherical surface (as in
Figs.~\ref{yyk_fig0}-\ref{yyk_fig1}), the $\tau_i$ steps are
usually unequal in size, unless a special effort is made to
partition a sphere into equal-area cells (see more in Kagan \&
Jackson 1998).
This cell inequality complicates the calculation.

Figs.~\ref{yyk_fig4} and \ref{yyk_fig5} show the EDs for both
Pacific regions.
The red curves are for the forecast, based on 1977-2003
seismicity, and the blue curves are for the earthquakes which
occurred in these regions from 2004-2006.
Both sets of curves are calculated using the forecast tables
like those in the example (Table~\ref{tab:Tab1}).
In principle, the calculations such as in (\ref{Eq_inf5a}) can
be made with unordered cells.
The density ordering in Table~\ref{tab:Tab1} and
Figs.~\ref{yyk_fig4},~\ref{yyk_fig5} is performed to create
the ED diagrams.

The score values $I_0$ in Table~\ref{tab:Tab2} are calculated
using the distribution shown by the red curves in
Figs.~\ref{yyk_fig4},~\ref{yyk_fig5}.
The $I_0$ values for NW- and SW-Pacific indicate that the
forecast yields an information score higher than 2-3 bits per
event.
This means that on average the probability gain (G) is a
factor of 5 to 10 ($2^{2.36}$ to $2^{3.38}$) when using the
long-term forecast compared to a random guess.
Of course, these $I_0$ values do not fully describe the
forecast advantage.
The boundaries of both regions have already been selected to
contain the maximum number of earthquakes in relatively small
areas.
If we extend any of the regions toward the seismically quiet
areas, the information score would significantly increase.
The proper measure of long-term forecast effectiveness
would extend the forecast method globally, i.e., over the
whole Earth surface.
Limited numerical experiments suggest that depending on degree
of smoothing, the value of $\epsilon$ (Eq.~\ref{Eq_inf-1}),
and other factors, the G-value for world-wide seismicity
varies from about 10 to 25.

The above values of the probability gain, G, can be compared
with similar calculations by Rhoades \& Evison (2005, 2006),
Console {\it et al.}\ (2006), and Rhoades (2007).
These authors calculated {\sl the information rate per
earthquake} for a model of smoothed seismicity (PPE), similar
to our long-term model.
The PPE model was compared to a stationary and spatially
uniform Poisson (SUP) model.
The probability gain, computed using the information rate, for
New Zealand, Japan, Greece, and California is about 4.5, 1.6,
1.6, and 3.4, respectively.
These relatively small gain values are related with the choice
by the authors of the regions that include only seismically
active areas (see {\it ibid.}).
Helmstetter {\it et al.}\ (2007, Table~1) obtained for
different long-term seismicity predictive models in California
the G-values ranging from 1.2 to 4.8.

\begin{figure}
\begin{center}
\includegraphics[width=0.5\textwidth,angle=0]{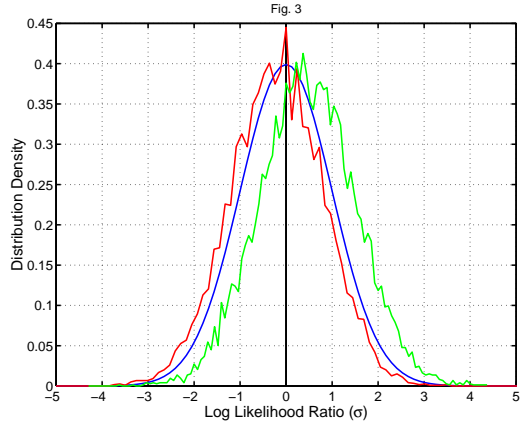}
\caption{\label{yyk_fig2}
Histograms of the log-likelihood function differences for
2004-2006 simulated earthquakes (see Fig.~\ref{yyk_fig1}).
The functions are normalized to have a unit standard
deviation.
We simulate 10,000 sets of 108 events for the NW-Pacific and
of 170 events for the SW-Pacific.
% We use the 1977-2003 earthquakes as a control set.
The blue line is the Gaussian curve with a zero mean and unit
standard deviation.
Red curve corresponds to simulation distributions for
NW-Pacific; green curve to SW-Pacific.
Curves on the right from the Gaussian curve correspond to
simulations that are worse than a real earthquake
distribution; curves on the left correspond to simulations
that are better than a real earthquake distribution.
}
\end{center}
\end{figure}

\begin{figure}
\begin{center}
\includegraphics[width=0.5\textwidth,angle=0]{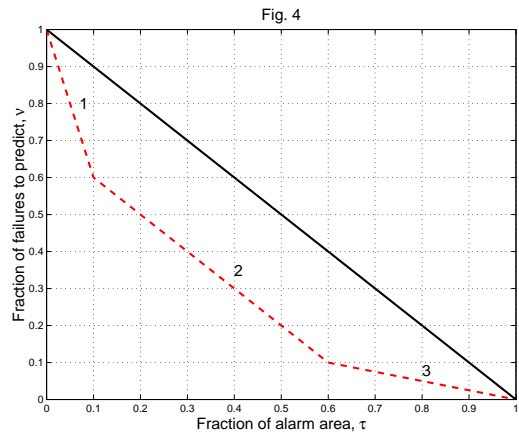}
\caption{\label{yyk_ede}
Error diagram $(\tau, \nu)$ example.
}
\end{center}
\end{figure}

\begin{figure}
\begin{center}
\includegraphics[width=0.5\textwidth,angle=0]{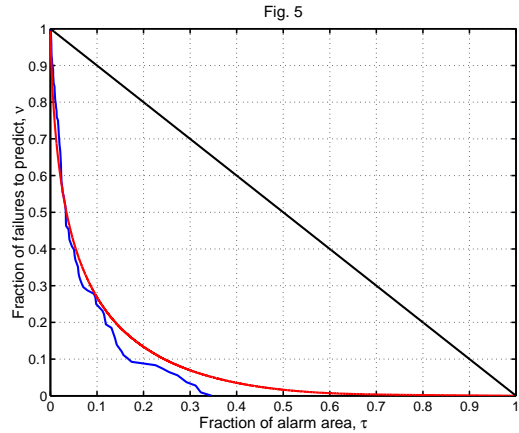}
\caption{\label{yyk_fig4}
Error diagram $(\tau, \nu)$ for NW-Pacific long-term
seismicity forecast.
The forecast is calculated at $121 \times 121$ grid.
Solid black line -- the strategy of random guess, red line --
the ordered density for long-term forecast, blue line --
earthquakes in 2004-2006.
}
\end{center}
\end{figure}

\begin{figure}
\begin{center}
\includegraphics[width=0.5\textwidth,angle=0]{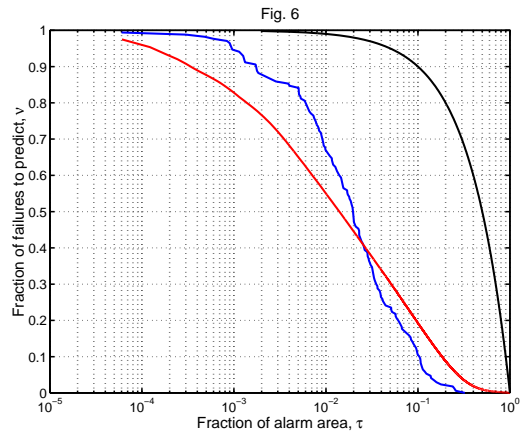}
\caption{\label{yyk_fig5}
Error diagram $(\tau, \nu)$ for SW-Pacific long-term
seismicity forecast: latitude
limits from $0.25^{\circ}$N to $60.25^{\circ}$S, longitude
limits from $109.75^{\circ}$E to $169.75^{\circ}$W.
The forecast is calculated at $121 \times 161$ grid.
Solid black line -- the strategy of random guess, red line --
the ordered density for long-term forecast, blue line --
earthquakes in 2004-2006.
}
\end{center}
\end{figure}

The ED curves for earthquakes in
Figs.~\ref{yyk_fig4},~\ref{yyk_fig5} are similar to the
forecast earthquake curves.
The computation of the likelihood scores (\ref{Eq_inf1},
\ref{Eq_inf2}) shows that the NW earthquakes have a better
score than the forecast, whereas SW events display the
opposite behavior (see also Fig.~\ref{yyk_fig2}).
The scores using the actual centroid position ($I_2$) are
larger than those for the cell centers ($I_1$), an anticipated
feature.
Similarly, Table~\ref{tab:Tab2} shows that the average scores
for synthetics ($<I_3>$) are very close to those of $I_0$,
which is understandable, since the simulation runs are
extensive (see Eqs.~\ref{Eq_inf3}, \ref{Eq_inf3a}).

\begin{figure}
\begin{center}
\includegraphics[width=0.5\textwidth,angle=0]{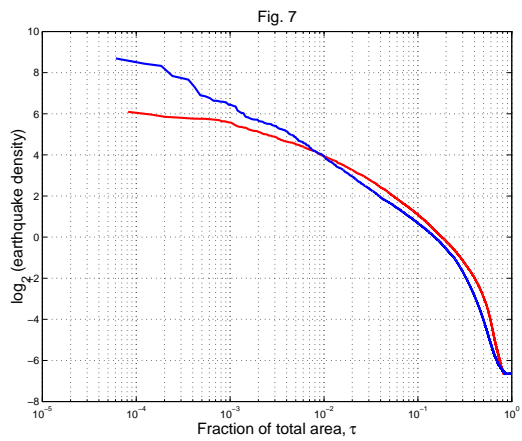}
\caption{\label{yyk_fig7}
% CONC_W2.M; WL2.PS
Likelihood function distribution for west Pacific long-term
seismicity forecasts: red -- NW-Pacific, blue --
SW-Pacific (see boundaries in
Figs.~\ref{yyk_fig0}, \ref{yyk_fig5}).
}
\end{center}
\end{figure}

Fig.~\ref{yyk_fig7} shows the frequency curves for the
log-likelihood function of both western Pacific regions.
We display $\log_2$ of the normalized rate (see column~5 of
Table~\ref{tab:Tab1}) against the normalized cumulative area
of the cells (column~4).
Curves for both regions exhibit high values of the rate ($R_i$)
concentrated in a relatively small fraction of area.
Low values at the right-hand end of the diagram correspond to
the assumed uniform background probability density
(Section~\ref{YYK_forecast}, see also
Figs.~\ref{yyk_fig0}-\ref{yyk_fig1}).

We calculate the higher order moments for the error curve
($I_0$ of Eq.~\ref{Eq_inf5} corresponds to the first moment
$\mu_1$)
\be
\mu_k \, = \, \sum_{i=1}^{N} \nu_i \,
\left [ \, \log_2 \left ( { { \nu_i }
\over {\tau_i } } \right ) - I_0 \, \right ]^k
\, ,
\label{Eq_inf5b}
\ee
where $k = 2,3,4,...$.

The standard deviation of the log-likelihood for the set of
$n$ events is
\be
\sigma_n \, = \, \sqrt { \mu_2 / n_2 }
\, ,
\label{Eq_inf5e}
\ee
where $n_2$ is the earthquake number during 2004-2006 (see
Table~ \ref{tab:Tab0}).
The coefficient of skewness is
\be
\eta \, = \, { \mu_3 / \mu_2^{3/2} }
\, ,
\label{Eq_inf5f}
\ee
and coefficient of kurtosis is
\be
\psi \, = \, { \mu_4 / \mu_2^2 } \, - \, 3
\, .
\label{Eq_inf5g}
\ee
These coefficients characterize how the likelihood curve
differs from the Gaussian distribution; for the latter law
both coefficients should be zero.
The {\sl Central Limit theorem} states that for large numbers
of i.i.d.\ events their
distribution should approach the Gaussian law.
If the event number is small, we need to find an efficient way
to numerically approximate the distribution of the sum of
i.i.d.\ random variables.

In Table~\ref{tab:Tab2} both coefficients are large for one
event likelihood curve (see also Fig.~\ref{yyk_fig7}),
but for the set of $n$ events they are small: the distribution
is close to the Gaussian law as demonstrated in
Fig.~\ref{yyk_fig2}.
The difference between the score values $I_0$ to $I_2$ is less
than the standard error value (see Table~\ref{tab:Tab2}).
Thus both forecasts can be considered statistically
successful.

The difference
\be
I^\prime \, =  \, I_0 - I_1 \quad {\rm or} \quad
I^{\prime \prime} \, =  \, I_0 - I_2
\, ,
\label{Eq_inf1b}
\ee
shows the predictive efficiency of a forecast, i.e., whether
on average earthquakes in 2004-2006 occurred at the sites
listed in the prediction table (see an example in
Table~\ref{tab:Tab1}).
For this particular time interval, both forecasts are
sufficiently good.
However, as other examples (Kagan \& Jackson 2000, Fig.~9;
Kagan {\it et al.}\ 2003, Fig.~5.2) demonstrate, this is not
always the case.
The values of differences (negative for the NW-Pacific and
positive for the SW-Pacific) correspond to those simulations
in Fig.~\ref{yyk_fig2}, where we display the distribution of
the difference $I_3 - I_2$.

By applying (\ref{Eq_inf5}) to the blue curve of earthquakes in
2004-2006 in Figs.~\ref{yyk_fig4},~\ref{yyk_fig5} we evaluate
the information score
\be
I_4 \, = \, { 1 \over n_2 } \sum_{i=1}^{n_2} \nu_i \,
\log_2 \left [ { { \nu_i }
\over {\tau_i } } \right ]
\, ,
\label{Eq_inf6}
\ee
(see Table~\ref{tab:Tab2}).
The value of $I_4$ is obviously significantly larger than all
the other estimates of the score.
Earthquake simulations provide an explanation for this
feature (see Fig.~\ref{yyk_fig12} below).

\subsection{Two-segment error diagrams and
information score }
\label{YYK_two }

\begin{figure}
\begin{center}
\includegraphics[width=0.65\textwidth,angle=0]{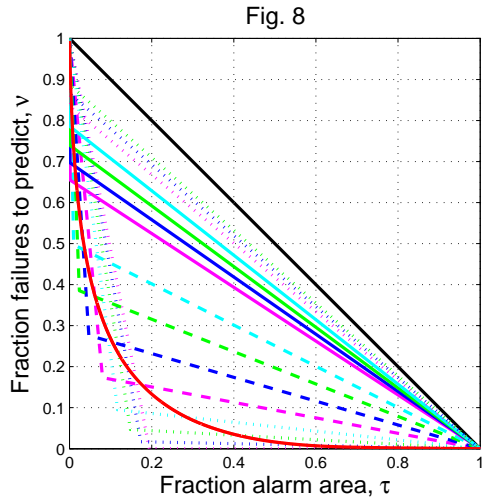}
\caption{\label{yyk_fig8}
Error diagram $(\tau, \nu)$ for NW-Pacific long-term
seismicity forecast, approximated by two-segment
distributions.
The solid thick black straight line corresponds to a random
guess, the thick red solid line is for the NW forecast.
Thin two-segment solid lines are for the curves with the
information score $I_0=2.3645$~bits.
The slope $D_1$ for the right-hand first segment is $D_1 =
- 2^{\, I_0} $.
For the next first segments slopes are
$D_1 \times 1.1$,
$D_1 \times 1.25$,
$D_1 \times 1.5$,
$D_1 \times 2.0$,
$D_1 \times 3.0$,
$D_1 \times 5.0$,
$D_1 \times 10$,
$D_1 \times 50$,
$D_1 \times 100$,
$D_1 \times 250$,
$D_1 \times 1000$,
$D_1 \times 10,000$,
$D_1 \times 100,000$,
and
$D_1 \times 1000,000$.
}
\end{center}
\end{figure}

Similarly to Fig.~5 in Kagan (2007b), in Fig.~\ref{yyk_fig8} we
display the approximation of the ED for the NW-Pacific by
several two line segment diagrams with the same value of
the information score, $I_0$.

For the assumed information score $I$, the contact point of two
segments is defined by the equation (corrected Eq.~22 by Kagan
2007b)
\be
D_1 \, \left [ \, { { \nu } \over { \, \nu - 1 - D_1 } }
\, \right ] ^{\nu } \, = \, - \, 2^{\, I}
\, .
\label{Eq_inf12}
\ee
By solving this equation for any value of the first segment
slope $D_1$ (non-positive by definition), one obtains the
$\nu$-value for the contact point of two linear segments,
$\tau~=~(\nu-1)/D_1$.

The first of these curves has the second segment coinciding
with the abscissa axis.
This means that one can obtain the same information score by
concentrating all the points in the $2^{-I_0} = 0.194$ part of
the region.
However, though the $I$-value for such a pattern would be
2.36~bits, all points would have the same value of the
probability gain.
Hence, for such a likelihood value distribution, the variance
and higher-order moments would be zero: very different from
the actual probability gain pattern (Table~\ref{tab:Tab1}).
If we modify the two-segment model to distribute the events
with different densities over the whole area, the variance and
the other moments would be non-zero.

\begin{figure}
\begin{center}
\includegraphics[width=0.65\textwidth,angle=0]{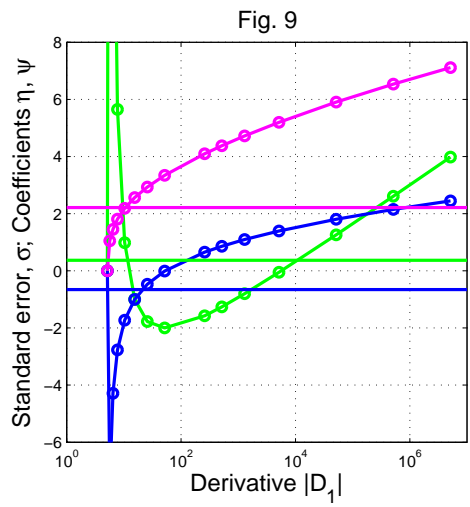}
\caption{\label{yyk_fig10}
Dependence on slope $|D_1|$ of standard deviation (magenta),
coefficients of skewness (blue) and kurtosis (green) for
two-segment curves in Fig.~\ref{yyk_fig8}.
Horizontal lines are these variables for the red curve in the
cited plot.
}
\end{center}
\end{figure}

In Fig.~\ref{yyk_fig10} we show the dependence of the
lower-order moments for the likelihood score on the $D_1$
slope.
For $D_1 = - 2 \times 2^I $ (dashed magenta line, fifth
curve from the bottom) the 2nd, 3rd, and 4th moments
correspond roughly to the moments of the forecasted densities.
Thus, such a two-segment model would reasonably well
approximate the actual event distribution.

The contact coordinates of two segments for this curve are:
$\nu_5 = 0.1732$ and $\tau_5 = 0.0803$.
Therefore, the point pattern having apparently the same
lower-order moments as the actual earthquake forecast would
have about 83\% of points concentrated in 8\% of the area,
i.e., the point density will be 10.3 times higher than the
uniform Poisson rate.
The rest of the events would be distributed in 92\% of the area
and have the rate of 0.19 compared to the uniform Poisson
distribution.
As we mention in Section~\ref{YYK_forecast}, in our Pacific
forecast 0.01 part of the total earthquake rate is spread over
the entire region (see Eq.~\ref{Eq_inf-1} and
Figs.~\ref{yyk_fig0}-\ref{yyk_fig1}).

\subsection{Information score for 1977-2003 CMT and PDE
catalogs }
\label{YYK_point4}

ED displays in Figs.~\ref{yyk_fig4},~\ref{yyk_fig5}
are inconvenient since the most interesting parts of the
curves are concentrated near $\nu-$ and $\tau$-axes.
The reason for this feature is that seismicity is concentrated
in relatively narrow seismic belts having a fractal spatial
earthquake distribution.
Now we focus on how other curves deviate from the forecasted
(red) one.
To make these deviations show more prominently, we need to
display the curves in a relative abscissa format, using the
1977-2003 forecast density as a template or baseline for the
likelihood score calculation.

\begin{figure}
\begin{center}
\includegraphics[width=0.5\textwidth,angle=0]{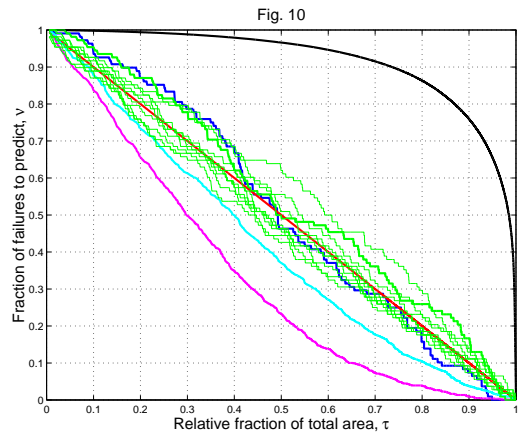}
\caption{\label{yyk_fig12}
Error diagram $(\tau, \nu)$ for NW-Pacific long-term
seismicity forecast.
Solid black line -- the strategy of random guess.
The solid thick red diagonal line is a curve for the NW
forecast.
Blue line is earthquake distribution from the CMT catalog in
2004-2006 (forecast);
magenta line corresponds to earthquake distribution from the
CMT catalog in 1977-2003;
cyan line is earthquake distribution from the PDE catalog in
1968-2006.
Thin green lines are ten simulations, displayed in
Fig.~\ref{yyk_fig1}, the first realization is shown by a thick
green line.
}
\end{center}
\end{figure}

Fig.~\ref{yyk_fig12} shows several curves in a new format; in
effect we convert the red curve in Fig.~\ref{yyk_fig4} to the
diagonal.
This is equivalent to calculating the information scores by
using $\lambda_i$ as a reference density
\be
{I_m} \, = \, { 1 \over n } \sum_{j=1}^n \nu_i \log_2 {
\zeta_i \over \lambda_i } \, ,
\label{Eq_inf1a}
\ee
where $\zeta_i$ is a rate density for all the other point
distributions.
Fig.~\ref{yyk_fig12} shows the difference between the forecast
curve (red) and the earthquake curve (blue) better than
Fig.~\ref {yyk_fig4}.

Fig.~\ref{yyk_fig12} also displays the curve for the
1977-2003 CMT catalog.
The numbers of events in the cell areas are shown in
Table~\ref{tab:Tab1}, column~8.
Also shown is the curve for the PDE catalog (U.S.\ Geological
Survey 2008) for 1968-2006.
We obtain $I_1~=~3.5991$ bits/event for the 1977-2003 CMT
catalog and $I_1~=~2.9789$ bits for the PDE.
These values are significantly larger than those forecasted
for 2004-2006.
Therefore, our forecast predicts better locations of past
earthquakes than those of future events.
Why this paradox?
In our forecast we use a broader smoothing kernel to capture
the spread of seismicity with time
(Section~\ref{YYK_forecast}).
Had we used the standard density estimation methods (Silverman
1986), the optimal kernel width would likely be smaller,
but such a smoothing would not effectively forecast future
seismicity.
A similar explanation is apparently valid for the PDE score
value.
Helmstetter {\it et al.}\ (2007, Table~1) obtained $G=7.1$
(significantly higher than the G-values for predictive
algorithms) when the same data were used to build the
long-term seismicity model and to test it (see
Section~\ref{YYK_point3}).
% Most likely, this feature of the forecast smoothing explains
% ``absurd post-diction" diagram by Kossobokov (2006, his
% Fig.~3), where the forecast applied in reverse

In Fig.~\ref{yyk_fig12} we also show several curves for
the simulated earthquakes.
These curves explain why the $I_4$-value (\ref{Eq_inf6}) is
significantly larger than the other measures of the
information score.
The reason is twofold.
First, the number of events in the 3-year interval is
relatively small and the curves often fluctuate around the
expected value (the red curve).
These fluctuations increase the sum value in (\ref{Eq_inf6}).
The curves are often below the red forecast line, which would
usually cause the score value to increase.
Second, the ED curve should be concave (Molchan 1997; 2003).
$I_4$-values, listed in Table~\ref{tab:Tab2}, are calculated
with the original curves shown in
Figs.~\ref{yyk_fig4},~\ref{yyk_fig5} which have many convex
sections.
If we make a lower envelope of the curve points, this would
decrease the $I_4$-value.
However, our numerical experiments show that the decrease is
not significant enough to bring the value sufficiently close
to the $I_0$ score.

The fluctuations of the synthetic curves also suggest that
some strategies proposed to measure the effectiveness
of a prediction algorithm by considering the ED, like a sum of
errors ($\nu+\tau$) or minimax errors (Molchan 1991; Molchan
\& Kagan 1992; Kossobokov 2006; Molchan \& Keilis-Borok 2008)
are biased for a small number of forecasted events.
For western Pacific regions the number of predicted events
($n_2$) is relatively large; in many other applications of the
ED ({\it ibid.}) this number is less than 10.

In Fig.~\ref{yyk_fig12} the forecast distribution curve is
used as the template.
Thus, we can measure the difference between this line and the
other curves using many standard statistical techniques, like
the Kolmogorov-Smirnov test, the Cramer-von Mises, etc.,
(Stephens 1974) to infer whether these distributions are
statistically different.

\section{Discussion }
\label{YYK_disc}

Several information scores are displayed in
Table~\ref{tab:Tab2}.
Although these scores appear different, the difference is
caused either by the small event number or a small number of
simulations.
The following limits can be easily conjectured
\be
I_0 \, =  \, \lim_{{\cal N}\to\infty} <I_3>
\, ,
\label{Eq_dis1}
\ee
(see Eq.~\ref{Eq_inf3a}).
In Table~\ref{tab:Tab2} the difference between these two
scores is small due to the large number of simulations.
Similarly,
\be
I \, =  \, \lim_{|S_i| \to 0} \, I_0 \, , \quad {\rm or} \quad
I \, =  \, \lim_{N \to \infty} \, I_0
\, ,
\label{Eq_dis2}
\ee
({\it cf.} Eq.~\ref{Eq_inf5}).
Also
\be
I_1 \, =  \, \lim_{|S_i| \to 0} \, I_2
\, ,
\label{Eq_dis3}
\ee
(see Eqs.~\ref{Eq_inf1}, \ref{Eq_inf2}).

In addition, if the model of the long-term forecast is
correct, then
\be
I_1 \, =  \, \lim_{n_2 \to \infty} \, I \, ,
\quad {\rm and} \quad
I_4 \, =  \, \lim_{n_2 \to \infty} \, I
\, ,
\label{Eq_dis4}
\ee
(see Eqs.~\ref{Eq_inf1}, \ref{Eq_inf6}).

In this paper we wanted to extend statistical analysis of
the stochastic point processes on line (usually time) to
multidimensional space.
In particular, we wished to find the relation between two
widely used statistical measures of prediction efficiency:
likelihood scores and error diagrams.
The equations derived here can be easily transformed to
describe quantitative connection between the information
scores and concentration diagrams (Section~\ref{YYK_intro}).

Summarizing our results, we list the following major points:
\hfil\break
$\bullet$
1. As with temporal stochastic processes (Kagan 2007b), we
find forward and inverse relations between the information
score and the error diagram curve for point spatial fields.
The error diagram represents a much more complete picture of
the stochastic point process than does likelihood analysis.
\hfil\break
$\bullet$
2. Since we are using a Poisson process to represent the
long-term spatial point pattern, the resulting models are
easier to visualize and calculate.
However, the assumption of earthquake statistical independence
and its influence on the information score value both need to
be investigated.
\hfil\break
$\bullet$
3. We extend our analysis for relatively small samples of
events and show that for such samples we should modify some of
the testing criteria proposed for error diagrams.
\hfil\break
$\bullet$
4. We show that the forecasting blueprint for estimating
future earthquake point density differs from standard
methods of statistical density evaluation.
Nevertheless, the connection between the likelihood score and
error diagrams described above can be used in many
density estimation problems.
\hfil\break
$\bullet$
5. We show that for testing the long-term forecast, it is
sufficient to process the forecast table to obtain the error
diagram and most information scores.
Thus, the simulation which was used in previous work, and which
requires significant computational resources, can be avoided
in most cases (Rhoades 2008).

\section*{Acknowledgments}
I appreciate partial support from the National Science
Foundation through grants EAR~04-09890, EAR-0711515, and
DMS-0306526, as well as from the Southern California
Earthquake Center (SCEC).
SCEC is funded by NSF Cooperative Agreement EAR-0106924 and
USGS Cooperative Agreement 02HQAG0008.
I thank David~D.~Jackson
of UCLA
and David Rhoades of GNS Science, New Zealand
for very useful discussions.
I thank Kathleen Jackson for significant improvements in the
text.
Publication 0000, SCEC.
%\end{acknowledgments}

% \clearpage
\newpage       %enable in DRAFT ??

% \newpage       %enable in DRAFT
% \end{article}  %enable in DRAFT

\newpage

\begin{table}
\begin{center}
\begin{tabular}{|r|r|c|r|c|c|r|r|}\hline
Lat. &  Long. & EQ Rate ($\lambda_i$) & $1^\circ \times 1^\circ $ & Cell Rate ($R_i$) & EQs & Probab. &  EQs \\
$\theta$ & $\phi$  & EQ/(day$\times$km$^2$) & ($S_i$) km$^2$ & EQ/[yr$\times(.5^\circ)^2$] & 04-06 & Gain (G) &  77-03 \\\hline
1 &  \multicolumn{1} {|c|} 2 & \multicolumn{1} {|c|} {3} & \multicolumn{1} {|c|} 4 & 5 & 6 & \multicolumn{1} {|c|} 7 &
\multicolumn{1} {|c|} 8 \\\hline
7.00 &  127.00 & 1.7909E-07 & 12254 & 0.2003970 & 0 & 68.3700 &  6 \\
40.50 & 143.00 & 1.6302E-07 &  9388 & 0.1397471 & 0 & 62.2331 &  1 \\
45.50 & 151.50 & 1.5134E-07 &  8653 & 0.1195765 & 0 & 57.7738 &  5 \\
24.00 & 122.00 & 1.4759E-07 & 11278 & 0.1519959 & 1 & 56.3445 & 10 \\
44.50 & 150.00 & 1.4496E-07 &  8805 & 0.1165462 & 0 & 55.3376 &  8 \\
44.50 & 149.50 & 1.4252E-07 &  8805 & 0.1145830 & 0 & 54.4055 &  9 \\
12.50 & 125.50 & 1.4152E-07 & 12053 & 0.1557541 & 0 & 54.0252 &  6 \\
44.00 & 148.50 & 1.4150E-07 &  8881 & 0.1147490 & 0 & 54.0181 &  8 \\
.... &.... &.... &....  &.... &.... &....  &....  \\
48.50 & 128.00 & 2.6198E-11 &  8180 & 0.0000196 & 0 &  0.0100 &  0 \\
48.50 & 127.50 & 2.6189E-11 &  8180 & 0.0000196 & 0 &  0.0100 &  0 \\
48.50 & 127.00 & 2.6183E-11 &  8180 & 0.0000196 & 0 &  0.0100 &  0 \\\hline
\end{tabular}
\caption{
Beginning and end of earthquake rate forecast table for
NW-Pacific based on CMT catalog (Ekstr\"om {\it et al.}\ 2005)
for 1977-2003 and ordered by descending rate density
(column~3, $\lambda$).
Cells are $0.5^\circ \times 0.5^\circ $, they form a $121
\times 121$ grid, EQ~-- earthquake(s).
% EQ~04-06~-- earthquake numbers in 2004-2006 (total 108),
% EQ~77-03~-- earthquake numbers in 1977-2003 (total 968).
}
\label{tab:Tab1}
\end{center}\end{table}

\newpage

\begin{table}
\begin{center}
\begin{tabular}{|c|c|c|c|}\hline
% P47, p.51
& & \multicolumn {2} {|c|} {Pacific Regions } \\ \cline {3-4}
& Time & NW & SW \\ \cline {3-4}
& Interval & \multicolumn {2} {|c|} {Annual Rate} \\\hline
$\upsilon_1$ &  77-03 Forecast & 35.7159 & 60.7509 \\
$n_1$ &  77-03 EQs & 968 & 1591 \\
$\upsilon_2$ &  77-03 EQs & 35.8546 & 58.9304 \\
$n_2$ &  04-06 EQs & 108 & 170 \\
$\upsilon_3$ &  04-06 EQs & 35.9918 & 56.6537 \\\hline
\end{tabular}
\caption{
Annual earthquake rate estimates.
Actual rate calculations are made with time intervals measured
in days (9861 days in 1977-2003 and 1096 days in 2004-2006).
For display convenience, we convert the daily rates into
annual rates by multiplying them by 365.25.
EQs~-- earthquakes.
}
\label{tab:Tab0}
\end{center}\end{table}

\newpage

\begin{table}
\begin{center}
\begin{tabular}{|r|c|r|r|}\hline
\# & Info  & \multicolumn {2} {|c|} {Pacific Regions ($n_2$)} \\ \cline {3-4}
   & Score & NW (108) & SW (170) \\\hline
% WTEST4A.FOR (P47,p.51)
1 & $I_0$ & 2.3645 & 3.3772 \\
2 & $I_1$ & 2.3675 & 3.0506 \\
3 & $I_2$ & 2.4067 & 3.2777 \\
4 & $<I_3>$ & 2.3609 & 3.3768 \\
5 & $I_4$ & 3.0970 & 3.9823 \\
6 & $I_0 - I_1$ & $-0.0030$ & 0.3266 \\
7 & $I_0 - I_2$ & $-0.0422$ & 0.1002 \\
 & & & \\
% WTESTPROB5.FOR (P47,p.52)
8 & $\sigma$ & 2.2102 & 2.9720 \\
9 & $\eta$ & $-0.6574 $ & $-0.2264$ \\
10 & $\psi$ & 0.3685 & $-0.3078$ \\
 & & & \\
11 & $\sigma_n$ & 0.2151 & 0.2296 \\
12 & $\eta_n$ & $-0.0196 $ & $0.0078$ \\
13 & $\psi_n$ & $-0.0369 $ & $0.0441$ \\\hline
\end{tabular}
\caption{Information scores for one event in
west Pacific regions, the standard error ($\sigma$) and
coefficients of skewness ($\eta$) and kurtosis ($\psi$)
(\ref{Eq_inf5e}-\ref{Eq_inf5g}).
The numbers in parentheses are event counts in
2004-2006 for each region, $n_2$.
Variables $\sigma_n$, $\eta_n$, and $\psi_n$ are for the set
of $n_2$ events.
}
\label{tab:Tab2}
\end{center}\end{table}

\newpage

%\end{article}   %disable in DRAFT

\end{document}